# Institutional Differences, Crisis Shocks, and Volatility Structure:

# A By-Window EGARCH/TGARCH Analysis of ASEAN Stock Markets

Author: Junlin Yang

Affiliation: Wenzhou-Kean University

Corresponding author: 3260700740@QQ.com

**Abstract**

This study investigates how institutional differences, and external crises jointly shape volatility dynamics in emerging Asian stock markets. Using daily stock index returns from Indonesia, Malaysia, and the Philippines from 2010 to 2024, the paper applies EGARCH (1,1) and TGARCH (1,1) models under a by-window estimation framework. The sample is divided into major global shock periods—the 2013 Taper Tantrum, the 2020–2021 COVID-19 pandemic, and the 2022–2023 global rate-hike cycle—along with their corresponding tranquil phases.

While GARCH-type models have been widely applied to single-market or static-period volatility studies, no existing research has systematically combined institutional comparison and multi-crisis dynamics within a unified modeling structure. This study directly addresses this methodological and empirical gap, offering a new framework for analyzing how institutional quality conditions the propagation and recovery of market volatility during global shocks.

The results reveal that all three markets exhibit strong volatility persistence ($\alpha + \beta \approx 1$) and fat-tailed return distributions ($v \in [5,8]$). During crises, volatility persistence and asymmetry ($\gamma$) rise significantly, while the degrees of freedom $v$ decline, indicating heavier tails and more frequent extreme fluctuations. Post-crisis periods show a gradual normalization of volatility structures. Cross-country comparisons highlight the buffering role of institutional maturity: Malaysia's well-developed regulatory and information systems dampen volatility amplification and accelerate recovery, while the Philippines' thinner market structure leads to prolonged instability.

The findings confirm that crises amplify volatility structures, whereas institutional robustness determines recovery speed. By bridging the disconnection between institutional economics and volatility modeling, this paper provides the first empirical evidence on the interactive mechanism of crisis amplification and institutional buffering in emerging markets. The results yield practical insights on enhancing market transparency, improving macroprudential communication, and strengthening liquidity support mechanisms to mitigate volatility persistence during global financial shocks.

**Keywords**

## 1. Introduction

Volatility plays a central role in assessing financial risk, market stability, and investor behavior. Since Engle (1982) introduced the Autoregressive Conditional Heteroskedasticity (ARCH) model, volatility modeling has become a cornerstone of empirical finance. Bollerslev (1986) generalized the framework to GARCH, enabling conditional variance to depend simultaneously on lagged innovations and past volatility—successfully capturing volatility clustering, one of the most persistent stylized facts of financial returns. Subsequent developments by Nelson (1991) and Glosten et al. (1993) introduced EGARCH and TGARCH models to account for asymmetrical market reactions to good and bad news, forming the foundation for modern volatility research.

Despite this extensive methodological development, existing studies have predominantly focused on **single-market or static-period analyses**. Empirical literature on emerging markets often treats institutional and crisis dimensions separately: some emphasize the persistence of volatility in developing economies (Choudhry, 1996; Li & Yu, 2020), while others explore crisis-induced volatility spikes without linking them to institutional resilience (Walid et al., 2011). This separation leaves a key empirical gap—**how institutional quality interacts with external shocks to shape volatility dynamics over time**. Addressing this intersection between institutional economics and time-varying volatility remains largely unexplored.

Emerging markets are particularly relevant for such inquiry. Their financial systems typically exhibit stronger volatility clustering, higher kurtosis, and slower mean reversion compared with developed markets (Tsay, 2005). At the same time, institutional characteristics—capital account openness, regulatory transparency, and macroprudential governance—differ substantially across countries (Chinn & Ito, 2008). These differences affect how markets absorb and transmit shocks during global crises such as the 2013 Taper Tantrum, the COVID-19 pandemic, and the 2022–2023 monetary tightening cycle. Therefore, understanding how institutional quality conditions volatility propagation and recovery is essential for both academics and policymakers.

This paper aims to fill that gap by combining institutional comparison and multi-crisis dynamics within a unified GARCH framework. Using daily data from Indonesia, Malaysia, and the Philippines (2010–2024), we estimate EGARCH and TGARCH models under a by-window approach that partitions the sample into crisis and post-crisis phases. This design allows parameters ($α, β, γ, ν$) to evolve across windows, revealing how volatility persistence, asymmetry, and tail risk respond to external shocks and how quickly markets revert to equilibrium afterward.

By doing so, the study contributes to the literature in three ways. First, it introduces a **dynamic institutional perspective** to volatility modeling, bridging a methodological divide between time-series econometrics and institutional economics. Second, it provides **cross-country evidence** on how structural differences in governance moderate crisis-induced volatility amplification. Third, it offers **policy-oriented insights** into how transparency, liquidity, and regulatory frameworks can mitigate volatility persistence in emerging markets. These contributions extend the frontier of GARCH-based research from static modeling toward institutional and systemic interpretation.

The remainder of this paper is organized as follows. Section 2 reviews prior studies on volatility modeling, institutional quality, and crisis dynamics. Section 3 presents the data and methodology, detailing the EGARCH and TGARCH models. Section 4 discusses empirical findings and robustness checks. Section 5 concludes with policy implications and directions for future research.

## 2. Literature Review

The theoretical and empirical study of volatility has evolved substantially since the early 1980s. Engle (1982) first introduced the Autoregressive Conditional Heteroskedasticity (ARCH) model to explain time-varying variance in macroeconomic and financial series. Bollerslev (1986) extended the framework to the Generalized ARCH (GARCH) model, allowing volatility to depend simultaneously on past squared shocks and lagged variances, thereby capturing volatility clustering observed in most financial assets. The subsequent introduction of asymmetric models—Exponential

GARCH (EGARCH) by Nelson (1991) and Threshold GARCH (TGARCH) by Glosten et al. (1993)—enabled researchers to capture the leverage effect, where negative shocks exert a stronger influence on volatility than positive ones. Together, these models laid the foundation for modern volatility analysis and risk measurement.

Beyond model structure, the distributional characteristics of financial returns have received significant attention. Empirical studies consistently find that return series are non-normal, exhibiting excess kurtosis and heavy tails (Tsay, 2005; Giller, 2005). To address this, researchers have incorporated alternative error distributions such as the Student-t and Generalized Error Distribution (GED), which better capture extreme observations and yield more robust parameter estimates (Eric, 2008). These innovations have broadened the applicability of GARCH models across asset classes, including equities, foreign exchange, and commodities.

Research on emerging markets provides a distinct perspective due to their structural and institutional heterogeneity. Choudhry (1996) and Li and Yu (2020) find that volatility in emerging markets is more persistent and less predictable than in developed economies, primarily because of limited market depth and information asymmetry. Chinn and Ito (2008) emphasize that institutional variables—such as capital account openness, regulatory transparency, and policy credibility—are critical determinants of market stability. Abdalla and Winker (2012) similarly demonstrate that stronger institutional frameworks moderate asymmetric volatility, highlighting that governance quality plays a direct role in financial risk transmission. These studies collectively suggest that institutional maturity may serve as a stabilizing force in volatility dynamics, but the interaction between institutions and crisis shocks remains underexplored.

The literature on financial crises has further expanded understanding of volatility behavior during systemic stress. Walid et al. (2011) use a Markov-switching EGARCH model to show that crises amplify volatility and strengthen co-movements between stock and exchange rate markets. Chen et al. (2022) adopt a by-window approach to trace parameter shifts across pre-crisis, crisis, and post-crisis periods, revealing significant structural changes in persistence and asymmetry. However, most studies either focus on one market or treat crises as isolated events rather than as recurring global phenomena. The dynamic interaction between crisis amplification and institutional buffering remains an open empirical question.

Building on these strands, the present study bridges the gap by combining institutional comparison and multi-crisis volatility modeling within a unified EGARCH/TGARCH framework. Unlike prior research that isolates institutional effects or examines single-event crises, this paper evaluates how institutional quality shapes both the amplification and the dissipation of volatility across multiple global shocks. In doing so, it introduces a comparative, multi-period perspective that connects financial econometrics with institutional economics, providing a richer understanding of volatility dynamics in emerging markets.

## 3. Methodology

To systematically capture the volatility characteristics of the three ASEAN stock markets and their dynamic behavior during crisis periods, this study adopts the Generalized Autoregressive Conditional Heteroskedasticity (GARCH) framework. Specifically, the **Exponential GARCH (EGARCH)** and **Threshold GARCH (TGARCH)** models are employed. Both models can capture the market's asymmetric reactions to positive and negative information and effectively describe the "bad-news amplification effect" commonly observed in financial time series.

### 3.1 Data sources and sample description

This study selects three ASEAN emerging markets—Indonesia, Malaysia, and the Philippines—as the research sample. The daily sample spans from January 2010 to December 2024. Stock index closing prices, USD/local-currency exchange rates, and the global volatility proxy (VIX) are retrieved from Yahoo Finance. Specifically, the stock index tickers are ^JKSE (Indonesia), ^KLSE (Malaysia), and PSEI.PS (the Philippines); the corresponding exchange rates are USDIDR=X, USDMYR=X, and PHP=X; and the global volatility index is ^VIX.

To analyze external shocks, the study follows the event timeline to define three crisis windows and the adjacent tranquil periods: the Taper period (May–December 2013), the COVID-19 period (March 2020–June 2021), and the rate-hike period (February 2022–June 2023). All data handling and model estimation are conducted in R, primarily using the rugarch, zoo, and tseries packages.

**3.2 Data preprocessing**

To ensure comparability across markets, raw prices are converted into log returns as:

$$r_t = \ln P_t - \ln P_{t-1}$$

and the first observation resulting from differencing is naturally missing.

A small number of missing points in stock and exchange-rate **price** series are linearly interpolated (`viana.approx( )`), after which returns are recomputed; the **VIX** series is aligned to trading dates using the same approach.

The crisis indicators **Crisis_Taper**, **Crisis_Covid**, and **Crisis_Hike** are constructed as *{0, 1}* dummies by the above time windows.

To verify the suitability of return series for GARCH-type modeling, the **Augmented Dickey–Fuller (ADF)** test is performed. The Dickey–Fuller statistics for Indonesia, Malaysia, and the Philippines are **–15.777**, **–15.686**, and **–15.729**, respectively, all with **p < 0.01**, rejecting the null of a unit root at the 1% level; hence all return series are stationary and appropriate for conditional heteroskedasticity models.

| Country | Test Statistic | Lag | p-value | Conclusion |
| --- | --- | --- | --- | --- |
| IDN | -15.777 | 15 | <0.01 | Stationary |
| MYS | -15.686 | 15 | <0.01 | Stationary |
| PHL | -15.729 | 15 | <0.01 | Stationary |

**Table 1**

The results of Table 1 further confirm the stationarity of the return series in all three countries, providing a solid statistical foundation for the subsequent EGARCH and TGARCH estimations.

**Model specification and parameter explanation**

After confirming the stationarity of the return series, this study further models the volatility structure using the Exponential GARCH (EGARCH) and Threshold GARCH (TGARCH) models. Both models belong to the generalized ARCH family and can effectively capture the dynamic dependence and asymmetry of volatility.

Let $r_t$ denote the stock return, decomposed as

$$r_t = \mu + \epsilon_t, \epsilon_t = \sigma_t z_t, z_t \sim t(0,1) \quad (1)$$

where μ is the conditional mean, $\epsilon_t$ is the innovation, $\sigma_t^2$ is the conditional variance, and $z_t$ follows a standardized **Student-t** distribution with degrees of freedom ν, accommodating fat tails.

For reference, the conventional GARCH (1,1) variance equation is

$$\sigma_t^2 = \omega + a\epsilon_{t-1}^2 + \beta\sigma_{t-1}^2 \quad (2)$$

where $\omega > 0$ is the long-run variance level, $a \geq 0$ captures the short-run impact of new shocks, and $\beta \geq 0$ characterizes volatility persistence; $a + \beta$ close to unity indicates high persistence.

To capture asymmetry (leverage effects), two models are employed:

**(a) EGARCH (1,1)** (Nelson, 1991):

$$\log(\sigma_t^2) = \omega + \beta\log(\sigma_{t-1}^2) \cdot a\frac{|\epsilon_{t-1}|}{\sigma_{t-1}} \cdot \gamma\frac{\epsilon_{t-1}}{\sigma_{t-1}} \quad (3)$$

where $\gamma$ measures asymmetric responses; $\gamma > 0$ implies that negative shocks amplify volatility more than positive shocks.

**(b) TGARCH (1,1) (Glosten, Jagannathan & Runkle, 1993):**

$$\sigma_t^2 = \omega + a\epsilon_{t-1}^2 + \gamma d_{t-1}\epsilon_{t-1}^2 + \beta\sigma_{t-1}^2 \quad (4)$$

where:

$$d_t = \begin{cases} 1, & \epsilon_{t-1} < 0 \\ 0, & \epsilon_{t-1} > 0 \end{cases}$$

Here $\gamma > 0$ indicates an additional volatility effect when returns are negative.

**Parameter meanings**: $\omega$ (baseline volatility); $a$ (reaction to new information); $\beta$ (volatility persistence); $\gamma$ (asymmetry / bad-news amplification); $v$ (tail thickness).

All models are estimated by **maximum likelihood (MLE)** under the Student-t assumption using **rugarch**. The **same specification and solver settings** are applied to all three markets to ensure cross-country comparability.

**3.4 Crisis sample division and subsample analysis**

To identify how volatility structure changes during and after crises, the full sample is divided into six windows: Pre-Taper (Jan 2010–Apr 2013), Taper (May–Dec 2013), Post-Taper (Jan 2014–Dec 2019), COVID-19 (Mar 2020–Jun 2021), Post-COVID (Jul 2021–Jan 2022), and Rate-hike (Feb 2022–Jun 2023). For each country and each window, EGARCH (1,1) and TGARCH (1,1) are estimated independently, and the key parameters ($\alpha, \beta, \gamma, \nu$) as well as the persistence indicator ($\alpha + \beta$) are summarized for comparison across phases.

| country | window | model | alpha | beta | gamma | nu | persistence |
|---|---|---|---|---|---|---|---|
| IDN | taper | EGARCH | -0.230230393 | 0.988522854 | -0.105578556 | 20.22243947 | 0.705503182 |
| IDN | taper | TGARCH | 2.03E-07 | 0.892561098 | 0.180174617 | 15.31149165 | 0.98264861 |
| IDN | post_taper | EGARCH | -0.078931153 | 0.979082739 | 0.105096025 | 5.311309879 | 0.952699598 |
| IDN | post_taper | TGARCH | 0.00381579 | 0.934689087 | 0.080644444 | 5.305119395 | 0.978827099 |
| IDN | covid | EGARCH | -0.094385046 | 0.933569796 | 0.299064334 | 6.764949263 | 0.988716917 |
| IDN | covid | TGARCH | 0.08754383 | 0.720965967 | 0.19193058 | 7.193139575 | 0.904475087 |
| IDN | post_covid | EGARCH | -0.087764699 | 0.643233504 | 0.05867083 | 8.782321026 | 0.58480422 |
| IDN | post_covid | TGARCH | 5.09E-05 | 0.999977603 | -0.008259868 | 10.28628041 | 0.995898615 |
| IDN | hike | EGARCH | -0.098479274 | 0.93595137 | 0.143133148 | 7.206711742 | 0.90903867 |
| IDN | hike | TGARCH | 1.46E-07 | 0.857711407 | 0.116714895 | 7.252942682 | 0.916069001 |
| MYS | taper | EGARCH | -0.310876543 | 0.893120227 | 0.167249361 | 6.023007302 | 0.665868364 |
| MYS | taper | TGARCH | 2.09E-06 | 0.740703089 | 0.376551128 | 5.091010308 | 0.928980739 |
| MYS | post_taper | EGARCH | -0.069426552 | 0.982580817 | 0.139364156 | 8.092032495 | 0.982836343 |
| MYS | post_taper | TGARCH | 0.0444755 | 0.911225137 | 0.058873965 | 7.883971345 | 0.985137619 |
| MYS | covid | EGARCH | -0.023806232 | 0.972174457 | 0.169120097 | 6.295112582 | 1.032928274 |
| MYS | covid | TGARCH | 0.037530096 | 0.922848547 | 0.029672365 | 6.413888239 | 0.975214826 |
| MYS | post_covid | EGARCH | 0.174572234 | 0.890741161 | -0.241150749 | 6.390082511 | 0.944738021 |
| MYS | post_covid | TGARCH | 0.001131858 | 0.999988605 | -0.091586584 | 8.566742337 | 0.955327171 |
| MYS | hike | EGARCH | -0.0763062 | 0.951029925 | 0.221549013 | 7.491741458 | 0.985498231 |
| MYS | hike | TGARCH | 0.063340229 | 0.837720566 | 0.131352838 | 7.067017401 | 0.966737214 |
| PHL | taper | EGARCH | -0.301273691 | 0.971998714 | -0.132759438 | 11.77322109 | 0.604345303 |
| PHL | taper | TGARCH | 8.72E-09 | 0.822884348 | 0.305960894 | 9.480673211 | 0.975864804 |
| PHL | post_taper | EGARCH | -0.062859666 | 0.988569047 | 0.02833235 | 10.37108648 | 0.939875555 |
| PHL | post_taper | TGARCH | 0.018295883 | 0.927474732 | 0.056479035 | 8.85941922 | 0.974010132 |
| PHL | covid | EGARCH | -0.062115008 | 0.949646322 | 0.324173076 | 3.601092507 | 1.049617853 |
| PHL | covid | TGARCH | 0.099674483 | 0.824505791 | 0.092787988 | 3.806488361 | 0.970574268 |
| PHL | post_covid | EGARCH | -0.29063507 | 0.945911212 | -0.226928304 | 14.70807787 | 0.54181199 |
| PHL | hike | EGARCH | -0.12654684 | 0.804817068 | 0.170274962 | 6.265340265 | 0.763407709 |
| PHL | hike | TGARCH | 1.54E-04 | 0.983013853 | 0.027015817 | 7.046872899 | 0.996676065 |

**Table 2**

The results in Table 2 indicate clear differences in volatility structure between crisis and tranquil periods across the three markets. In general, during crisis windows, α + β rises markedly—indicating stronger volatility persistence—while the asymmetry parameter γ increases across the board, reflecting intensified bad-news amplification. At the same time, the degrees of freedom ν are lower than in tranquil periods, implying heavier tails in the return distribution and a higher probability of extreme fluctuations. Taken together, these features reveal the amplification effect of crisis shocks on market risk structure.

At the country level, Indonesia's α + β approaches unity and γ rises significantly during the Taper and COVID phases, indicating that both the persistence and the asymmetry of volatility shocks intensify. As the crises recede, α + β falls to around 0.6 in the post-COVID phase, suggesting a relatively fast reversion of volatility. This crisis-to-recovery parameter trajectory is consistent with the evidence reported by Choudhry (1996) for Asian markets, where improvements in capital flows and policy stability are typically accompanied by a pronounced decline in volatility persistence.

Malaysia exhibits an increase in volatility during crises but remains overall stable. EGARCH estimates for the COVID phase show α + β slightly above 1, implying near unit-root persistence; however, this effect fades quickly in the post-pandemic phase, with γ turning negative, indicating that good news reduces volatility more than bad news raises it. This "peak-and-rapid-reversion" pattern aligns with Walid et al. (2011), who argue that transparent information mechanisms and macroprudential frameworks help accelerate volatility normalization in more robust institutional environments—consistent with Malaysia's relatively well-developed regulatory system and disclosure practices.

The Philippines displays the most dramatic parameter shifts. During COVID, α + β exceeds 1 while γ reaches 0.32, implying near-explosive volatility under crisis conditions. Although volatility declines after the crisis, the speed of reversion is evidently slower than in Indonesia and Malaysia. This is broadly in line with Li and Yu (2020), who find that markets with thinner liquidity and weaker institutional constraints tend to exhibit stronger asymmetry and persistence under stress. The evidence suggests that the Philippine market's relatively limited depth and narrow investor base make high volatility more persistent in the face of external shocks.

Overall, the subsample estimates corroborate the amplification of volatility structure during crises and the buffering role of institutional differences. Markets with greater institutional maturity recover more quickly after crises, whereas markets with weaker institutions exhibit higher persistence and stronger bad-news sensitivity.

*Model diagnostics and robustness tests*

To verify the reliability of model fit and the robustness of parameter estimates, this study conducts multiple diagnostics based on the standardized residuals, including tests for serial correlation and remaining heteroskedasticity, as well as robustness checks for the error distribution.

First, standardized residuals from the EGARCH models are examined visually for all three countries.

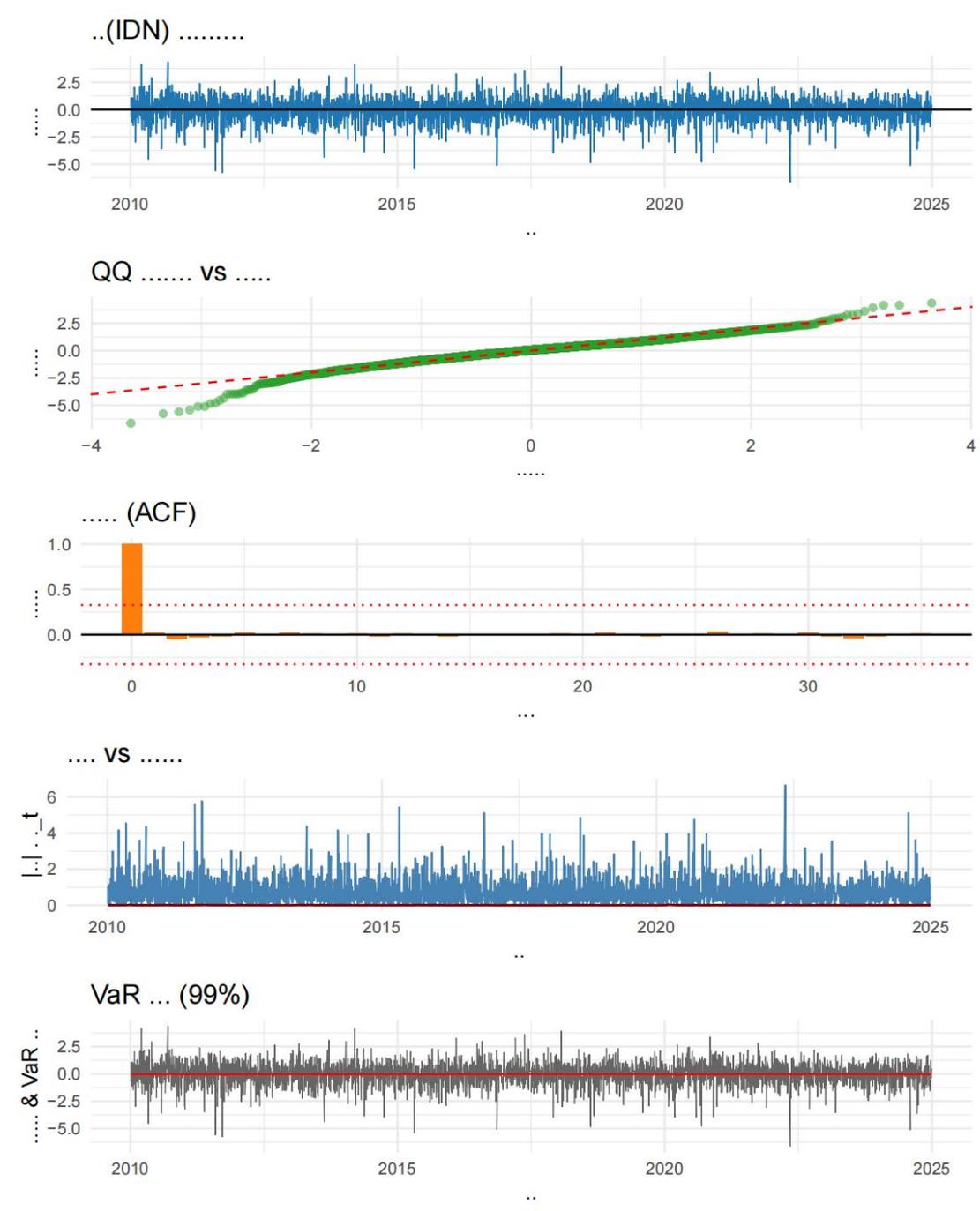

**Graph 1**

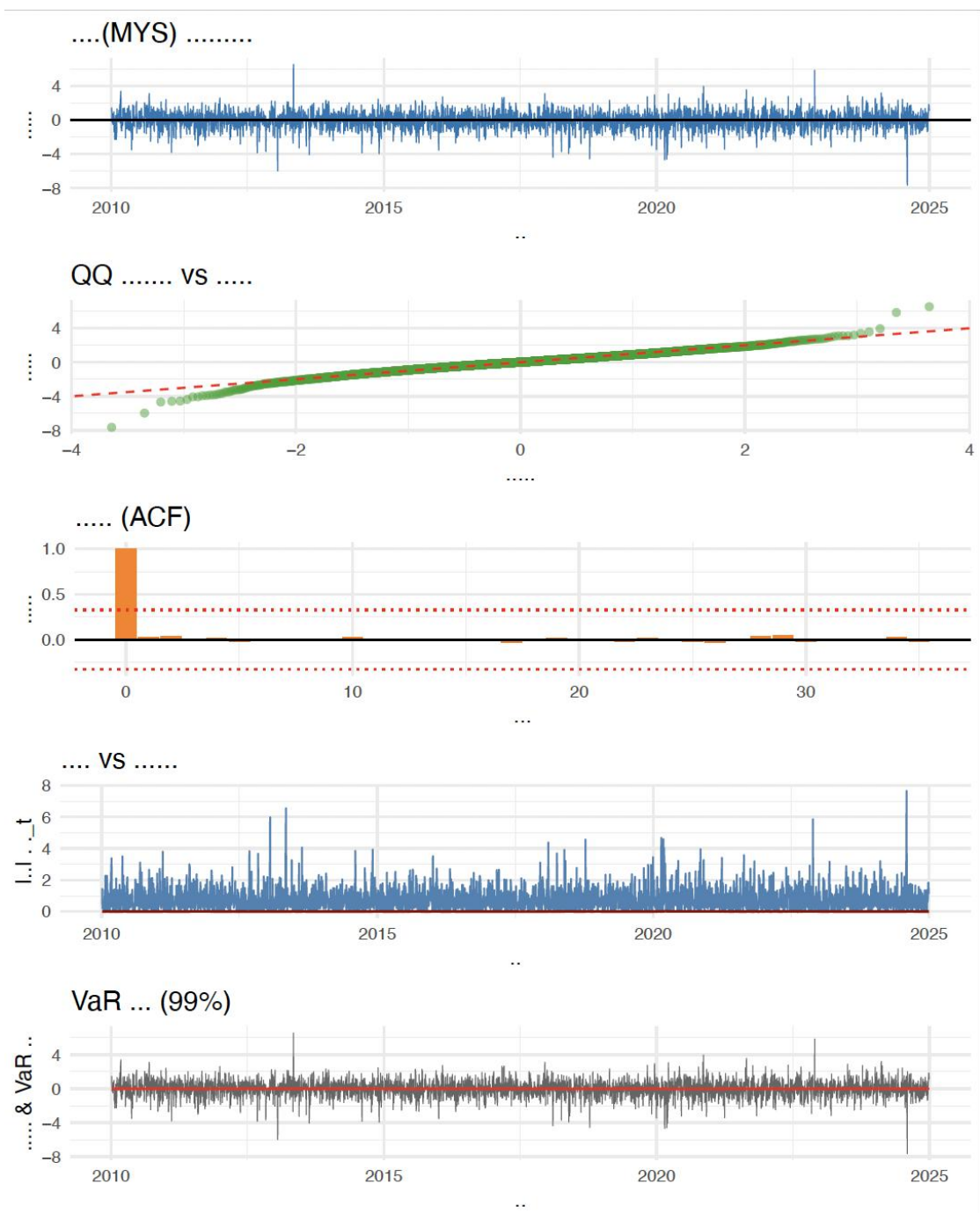

**Graph 2**

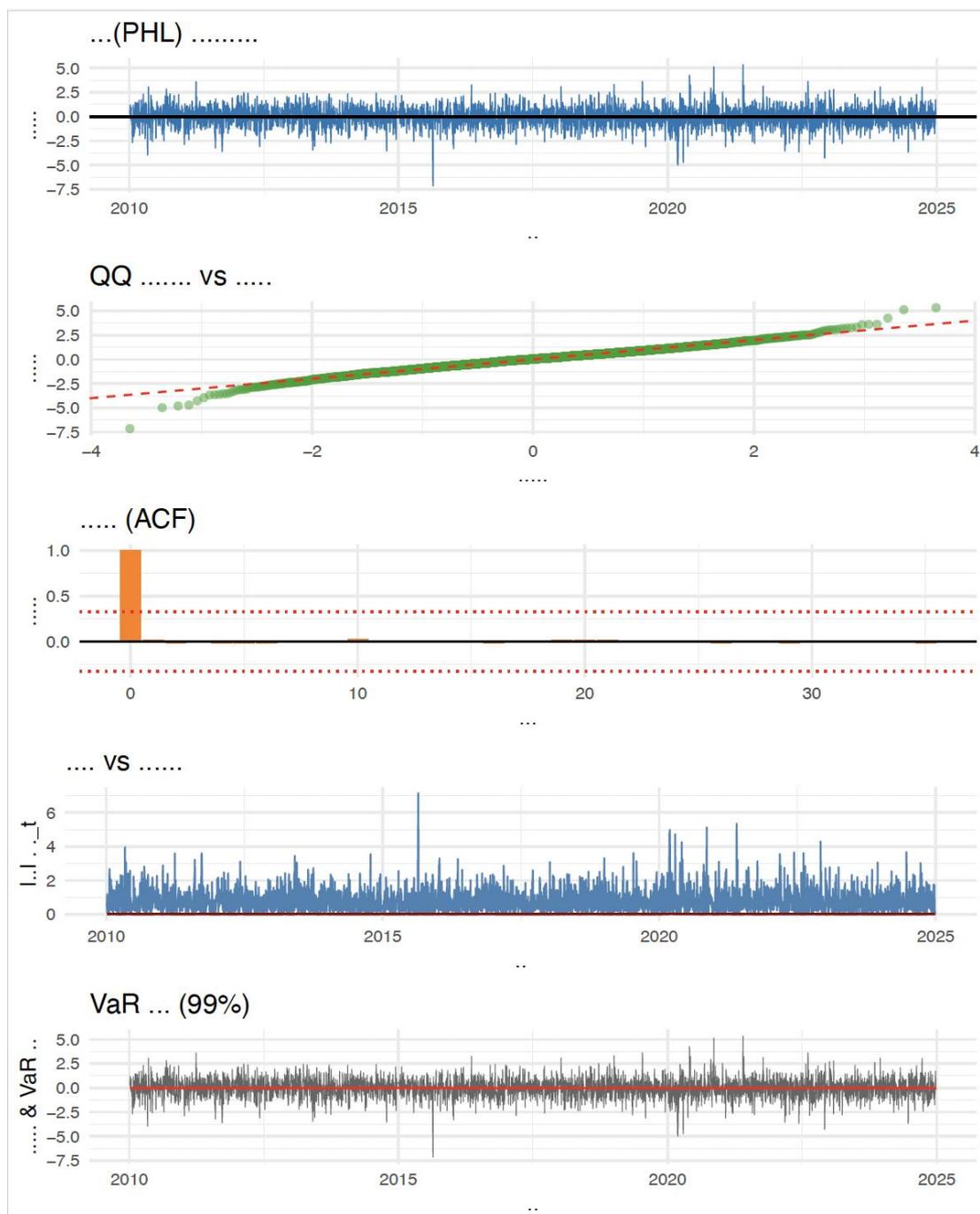

**Graph 3**

From Figures 1 to 3, the standardized residuals fluctuate around zero with no apparent trend. The residual ACFs show that most autocorrelations lie within the 95% confidence bands, indicating that the mean equation has adequately captured linear structure. The QQ plots align closely with the theoretical line, with minor tail deviations consistent with the heavy-tailed t-distribution assumption. The model-implied volatility tracks the realized |residual| dynamics well, suggesting that the models follow volatility clustering in real time. The VaR hit plots show that most observations remain within the 99% risk threshold, with only a few exceedances early in crisis phases—evidence that the risk measures are generally reasonable.

To further assess statistical adequacy, Ljung–Box and ARCH–LM tests are applied to the standardized residuals. The results show that Indonesia's Ljung–Box p-value is 0.045 while its ARCH–LM p-value is 0.942, indicating residuals close to white noise and no remaining conditional heteroskedasticity. For Malaysia, the p-values are 0.099 and 0.020; for the Philippines, 0.593 and 0.030. These imply no significant serial correlation and only mild residual ARCH at the 5% level. Overall, the residual properties conform to the assumptions of the GARCH family, supporting the reliability of the fitted models.

| Country | Ljung–Box p-value | ARCH–LM p-value | Interpretation |
| --- | --- | --- | --- |
| Indonesia (IDN) | 0.045 | 0.942 | Residuals are nearly white noise; no remaining heteroskedasticity. |
| Malaysia (MYS) | 0.099 | 0.020 | No significant autocorrelation; slight remaining ARCH effect. |
| Philippines (PHL) | 0.593 | 0.030 | No significant autocorrelation; mild remaining ARCH effect. |

**Table 3**

Additionally, to test robustness to the error distribution, the Student-t specification is replaced by the generalized error distribution (GED). The GED shape parameters for Indonesia, Malaysia, and the Philippines are 1.349, 1.350, and 1.446, all below 2, indicating that the return distributions remain heavy-tailed. The corresponding log-likelihood values—12309.61, 13887.12, and 12127.87—are very close to those under the t-distribution, suggesting that the main estimates and inferences are stable.

| Country | Shape Parameter | Log-Likelihood | Interpretation |
|---|---|---|---|
| Indonesia (IDN) | 1.349 | 12309.61 | Thick tails (shape < 2); model remains stable under GED. |
| Malaysia (MYS) | 1.350 | 13887.12 | Similar to Indonesia; thick-tailed distribution confirmed. |
| Philippines (PHL) | 1.446 | 12127.87 | Slightly thinner tails but still far from normality; robustness verified. |

**Table 4**

In sum, the diagnostics and robustness checks show that the models fit well, the distributional assumption is reasonable, and the main inferences are not sensitive to the error distribution. Both EGARCH and TGARCH effectively characterize volatility dynamics and asymmetry in the three markets, providing a solid foundation for subsequent crisis comparisons and institutional analysis.

## *Comprehensive analysis*

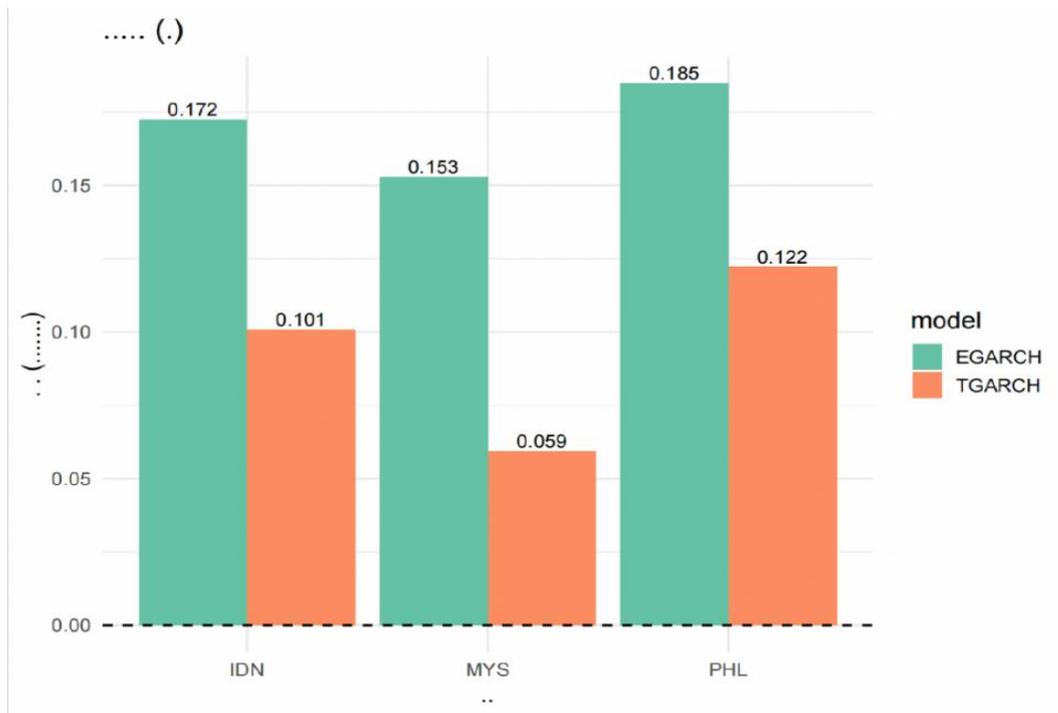

**Graph 4**

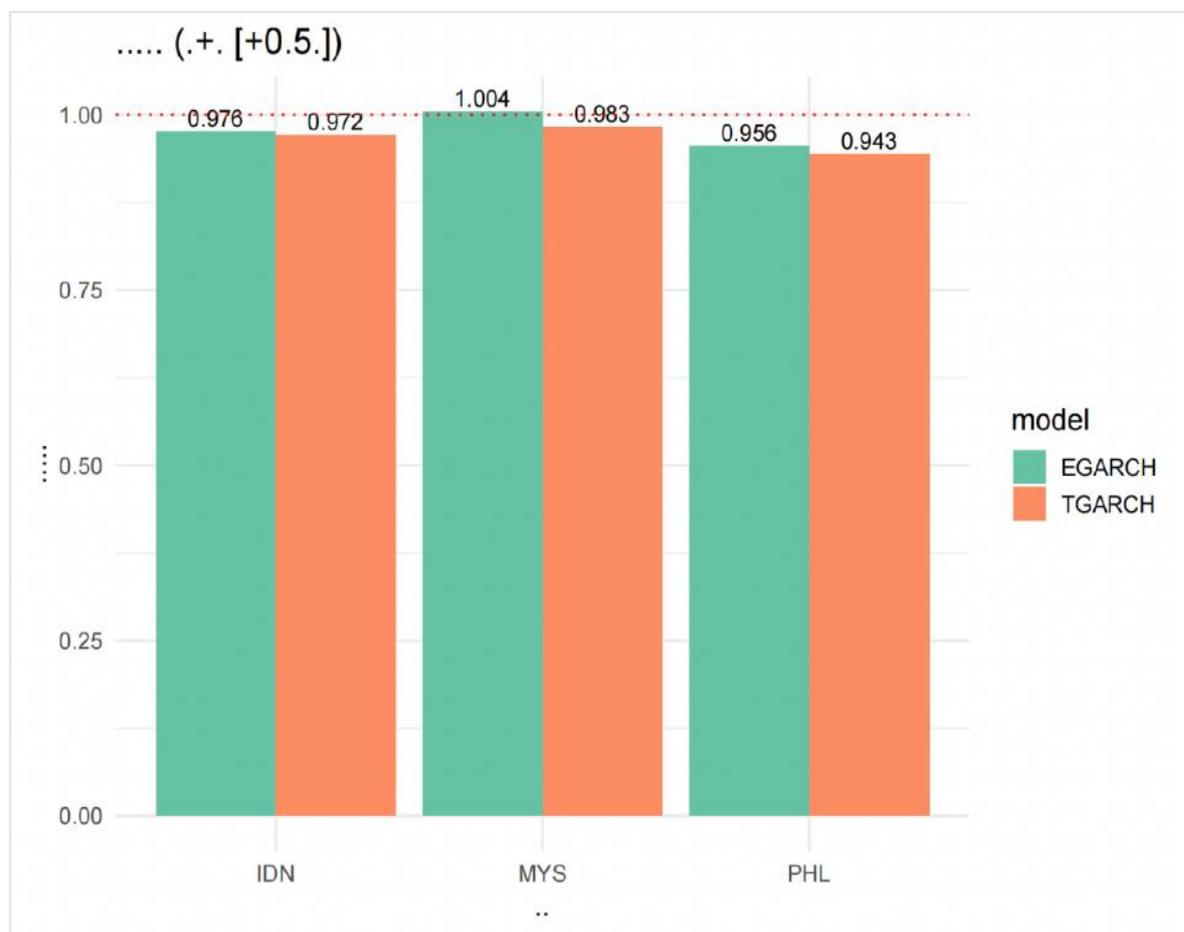

**Graph 5**

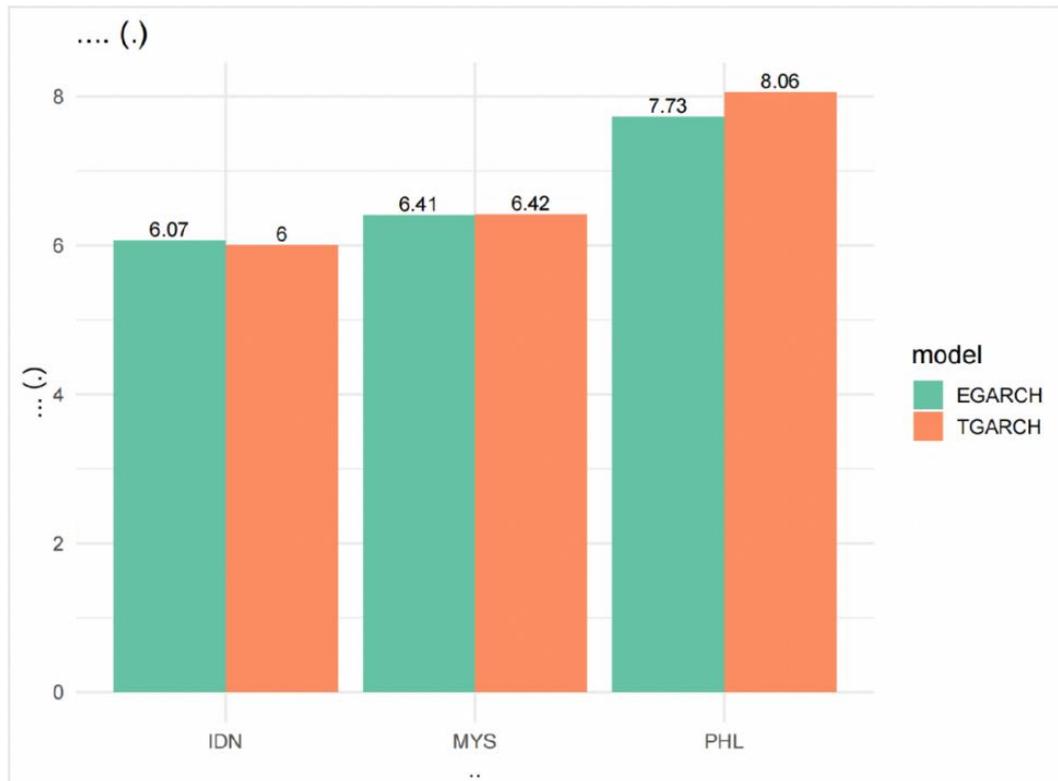

**Graph 6**

From the cross-sectional comparison in Figures 4–6, the Philippines records the highest asymmetry parameter γ, followed by Indonesia, with Malaysia the lowest. The α + β indicator is close to unity in all three markets, indicating pervasive persistence. The degrees of freedom ν mostly fall within the 5–8 range, confirming pronounced fat tails. Consistent with Abdalla and Winker (2012) and Tsay (2005), markets with stronger institutions exhibit more "tempered" behavior in both \gamma and \nu, which provides a benchmark for the parameter transitions observed in crisis windows.

Synthesizing the empirical evidence, the three ASEAN markets share common features of emerging economies while reflecting an interaction between institutional differences and crisis shocks. Overall, all three markets display significant volatility persistence and fat-tailed returns; asymmetry strengthens notably during crises; and markets with greater institutional robustness tend to recover faster thereafter.

From the model estimates, $α + β$ near unity indicates long-lasting volatility. Crises push this indicator even higher, implying that uncertainty amplifies the duration of shocks. At the same time, $γ$ increases significantly in crisis windows—most prominently in the Philippines—suggesting that negative information is further magnified by investor sentiment and liquidity constraints. By contrast, Malaysia's $γ$ declines swiftly after crises and can even turn negative, reflecting stronger resilience under a more developed institutional and liquidity environment.

Regarding the return distribution, $\nu$ between 5 and 8 confirms fat tails; $\nu$ declines further during crises, indicating an increased likelihood of extreme moves. This pattern is corroborated by the GED robustness checks, where shape parameters around 1.3–1.4, far below the normal-equivalent value of 2, confirm that tail thickness is a data feature rather than an artifact of the chosen distribution.

Residual diagnostic results also point to adequate model fit. Indonesia's residuals are essentially white noise with no remaining heteroskedasticity; Malaysia and the Philippines exhibit only minor residual ARCH, which does not affect overall validity. The diagnostic figures show that model-implied volatility tracks realized fluctuations closely, indicating that EGARCH and TGARCH capture time-varying risk effectively.

Taken together, the evidence supports the following conclusions: crises intensify volatility persistence and bad-news amplification, while institutional robustness governs the speed and extent of post-crisis normalization. In particular, Malaysia's stronger institutions act as a "buffer," enabling faster reversion to equilibrium; Indonesia lies in the middle; and the Philippines, with a thinner market structure, experiences more severe and prolonged volatility. Hence institutional differences shape not only the level of volatility but also market fragility and self-repair capacity during crises.

*Conclusions and mechanism interpretation*

This paper conducts unified EGARCH and TGARCH modeling and by-window estimation for Indonesia, Malaysia, and the Philippines using daily data from 2010 to 2024, covering the Taper, COVID, and rate-hike crises. The conclusions can be summarized along three lines—common features, crisis effects, and institutional moderation.

First, common features. ADF tests confirm stationarity of the three-return series. In full-sample estimation, α + β is generally close to one, indicating high persistence typical of emerging markets. The Student-t degrees of freedom ν concentrate in the 5–8 range, and the GED shape parameters cluster around 1.3–1.4, jointly confirming fat tails. Together, these constitute the baseline: persistent volatility and a non-negligible probability of extremes throughout the sample.

Second, crisis effects (see Table 2). During crisis windows, α + β rises further, γ strengthens significantly, and \nu declines—corresponding to "stickier, more asymmetric, and heavier-tailed" dynamics. Specifically, uncertainty keeps shocks in conditional variance longer (persistence transition); negative information has a larger marginal impact (bad-news amplification); and thicker tails raise the likelihood of extremes. After crises, parameters revert, indicating a transition from exogenous shock dominance to endogenous repair.

Third, institutional moderation. Cross-sectionally, Malaysia's γ is lower and reverts faster, Indonesia lies in the middle, and the Philippines is highest with slower reversion. This implies that stronger institutions suppress the amplification of asymmetry and persistence during crises and accelerate recovery. Put differently, institutions serve as a "buffer" in shock transmission: transparent disclosure, predictable macroprudential frameworks, and deeper liquidity help reduce the peak of $\gamma$ and shorten the high-plateau duration of $α + β$, thereby compressing the length of high-volatility periods.

Diagnostics and robustness checks support these conclusions: standardized residuals show no material serial correlation by Ljung–Box; ARCH–LM indicates no remaining heteroskedasticity for Indonesia and only mild remnants for Malaysia and the Philippines; GED re-estimation yields stable likelihoods and parameter signs, demonstrating that the main findings are not driven by the error distribution assumption. Overall, the evidence shows: crises amplify the volatility structure, while institutions determine the recovery path. These conclusions are mutually reinforced by the figures and tests reported in the main analysis.

*Policy implications and implementation pathways*

Based on the "crisis amplification—institutional buffering—accelerated recovery" evidence, policy should focus on two main levers—information and liquidity—and a country-differentiated schedule, with an emphasis on operational feasibility and sequencing.

1. Information lever: lower the crisis peak of $\gamma$.

In the early crisis stage, supervisors should adopt forward-looking communication and standardized risk messaging, with clear, scheduled, and auditable disclosures on key items (e.g., temporary trading rules, valuation and settlement arrangements, activation of macroprudential tools). Transparent and continuous communication dampens the chain from bad news to expectation spirals to volatility amplification, suppressing the peak and inertia of γ. Issuer and intermediary disclosure guidelines should be coordinated so that risk information reaches investors quickly and uniformly.

2. Liquidity lever: shorten the high-plateau duration of α + β.

During stress, countercyclical margins, market-maker inventory exemptions, temporary re-pledge/liquidity facilities, and similar tools can directly alleviate trading frictions and reduce the "dwell time" of volatility in conditional variance. These tools should be equipped with automatic triggers and exit thresholds to prevent policy uncertainty from becoming a new volatility source. Coordination between cash and derivatives markets (e.g., during abnormal index futures discounts) can further reduce spillovers.

3. Country-differentiated implementation: target-setting, sequencing, and sizing.

    Malaysia: robust institutions and fast recovery—prioritize transparency and predictability, minimizing the frequency and magnitude of policy shifts to avoid secondary shocks to γ.

> Indonesia: medium recovery—seek a better balance between capital-flow management and trading transparency; broaden the liquidity toolkit (market-making, ETFs) to compress the high-plateau duration of α + β.

The Philippines: strongest shock, slowest recovery—first increase market depth (market-maker incentives, auction efficiency) and adopt a predictable policy framework (communication cadence, disclosure templates) to reduce the elasticity of γ at the source and curb tail risk exposure.

4. Sequencing: a three-stage schedule.

Early (shock identification): emphasize the information lever to stabilize expectations and lower the peak of γ;

Middle (liquidity repair): activate the liquidity lever to shorten the high-plateau stage of α + β;

Late (reversion and codification): evaluate and calibrate, codify lessons, and improve the "automatic stabilizer" design of macroprudential and disclosure rules.

In short, policy effectiveness lies not in the mere presence of measures but in their targeted adjustment of γ and α + β: reducing the marginal amplification of bad news and shortening the residence time of high volatility is the key to shifting volatility from a "shock-driven" to a "mechanism-driven" regime.

*Limitations and future research*

The conclusions rest on three testable premises: (i) stationarity of returns verified by ADF tests; (ii) residual properties of EGARCH/TGARCH broadly adequate per Ljung–Box and ARCH–LM, with only mild residual heteroskedasticity in a few cases; and (iii) stability to the error distribution verified by GED robustness. Subject to data and scope, three limitations remain:

(1) No exogenous regressors in the variance equation.

This study obtains stable results without exogenous terms but does not include VIX, exchange-rate volatility, etc., directly in the variance equation. Future work can test GARCH-X or EGARCH with exogenous regressors.

(2) Limited asset coverage.

The focus is on equity index returns only, excluding bonds and FX. Future research can extend to multi-asset estimation and use multivariate GARCH (e.g., DCC/BEKK) to identify cross-market spillovers.

(3) Qualitative treatment of institutions.

Institutional differences are identified via cross-country comparison, but no fine-grained institutional index is constructed. Future work can incorporate regulatory transparency, capital-account openness, rule-of-law and governance indices to quantify how institutions modulate γ and α ≠ β.

Further research could combine event-study designs with high- or ultra-high-frequency data to identify shock transmission channels and microstructure mechanisms; semi-parametric or machine-learning approaches could also be considered to better capture tail risk and outliers, thereby balancing predictive accuracy and interpretability.